\documentclass[aps,twocolumn]{revtex4}
\usepackage{epsfig,amsbsy,bm}
\newcommand{\isum}%
{\mathop{\hbox{$\displaystyle\sum\kern-13.2pt\int\kern1.5pt$}}}

\renewcommand{\k}{{\bm k}}
  \newcommand{\n}{\hat {\bm k}}
  \newcommand{\p}{{\bm p}}

\newcommand{\eref}[1] {(\ref{#1})}
\newcommand{\Eref}[1] {Eq.~(\ref{#1})}
\newcommand{\Fref}[1] {Figure \ref{#1}}
\newcommand{\br}{\begin{eqnarray*}}
\newcommand{\er}{\end{eqnarray*}}
  
\newcommand{\be}{\begin{equation}}
\newcommand{\ee}{\end{equation}}
\newcommand{\hs}{\hspace*}
\newcommand{\vs}{\vspace*}
\newcommand{\e}{\hat {\bm e}}
\newcommand{\ba}{\begin{eqnarray}}
\newcommand{\ea}{\end{eqnarray}}

\begin{document}
\bibliographystyle{prsty}
\voffset=2cm

\title
{Different escape modes in   two-photon
double ionization of helium}

\author{A. S. Kheifets} \protect\email{A.Kheifets@anu.edu.au} 
\author{A. I. Ivanov}

\affiliation{Research School of Physical Sciences,
The Australian National University,
Canberra ACT 0200, Australia}

\author{I. Bray}
\email{I.Bray@murdoch.edu.au}
\affiliation{ARC Centre for Matter-Antimatter Studies,
         Murdoch University, Perth, 6150 Australia
}

\date{\today}
\begin{abstract}
The quadrupole channel of two-photon double ionization of He exhibits
two distinctly different modes of correlated motion of the
photoelectron pair.  The mode associated with the center-of-mass
motion favors a large total momentum which is maximazed at a parallel
emission. However, the mode associated with the relative
motion favors an antiparallel emission. This difference is manifested
in a profoundly different width of the angular correlation functions
corresponding to the center-of-mass and relative motion modes.

\end{abstract}

\pacs{32.80.Rm 32.80.Fb 42.50.Hz}
\maketitle

The process of correlated motion of multiple ionization fragments has
been at the forefront of atomic collision physics during the past
decade. Recent progress in experimental techniques made it possible to
detect simultaneously a large number of charged reaction fragments
with fully determined kinematics \cite{URD03}. The long range Coulomb
interaction between these fragments makes a full theoretical
description of such a process a highly challenging task.
In the meantime, the simplest multiple fragmentation reaction, the
single-photon double ionization (SPDI) of helium is now well
understood with accurate theoretical predictions being confirmed
experimentally under a wide range of kinematical conditions
\cite{BS00,KA00,AH05}.
All the information about the correlated motion of the photoelectrons
is described in SPDI by a pair of symmetrized amplitudes
$f^\pm(\theta_{12},E_1,E_2)$ which depend on the relative interelectron
angle and energy \cite{HSWM91,MSH97}. If the photon is described by
the linear polarization vector $\e$, the dipole matrix element
of SPDI can be written simply as
\be
\label{dipole}
D \propto
f^+ \ (\n_1+\n_2)\cdot \e 
+
f^- \ (\n_1-\n_2)\cdot \e 
\ ,
\ee
where $\n_i=\k_i/k_i$, $i=1,2$ are the unit vectors directed along the
photoelectron momenta $\k_i$. Generalization of \Eref{dipole} to
arbitrary polarization of light is straightforward \cite{S95}.  Under
the equal energy sharing condition, the anti-symmetric amplitude
vanishes $f^-(E_1=E_2)=0$ and all the information about the SPDI
process is contained in one symmetric amplitude $f^+$.  Following
predictions of the Wannier-type theories
\cite{Rau76,Feagin84}, the SDPI amplitude can be written using the
Gaussian ansatz:
\be
\label{gauss}
 |f^+|^2  \propto
\exp\left[ -4\ln2{(\pi-\theta_{12})^2\over\Delta\theta_{12}^2}
\right]
\ee
where the width parameter  $\Delta\theta_{12}$ indicates the strength
of angular correlation in the two-electron continuum. Although the
analytical theories \cite{Rau76,Feagin84} validate \Eref{gauss} only
near the double ionization threshold, numerical models \cite{KB02} and
direct measurements \cite{KBB03,KKKB05}  support its
validity in a far wider photon energy range \cite{KB02}.

Two-photon double ionization (TPDI) of He is a much more complex
fragmentation process with two competing decay channels into the $S$
and $D$ two-electron continua. In a most general case, complete
separation of the dynamic and kinematical variables in TPDI requires
introduction of five symmetrized amplitudes \cite{KI06} or four
unsymmetrized amplitudes \cite{DAMOP06}.

In this Letter, we demonstrate that complexity of the TPDI leads to a
new phenomenon of two distinct correlated escape modes: one is
associated with the center-of-mass motion of the photoelectron pair
whereas another is related to their relative motion. The
center-of-mass motion favors a large total momentum of the pair
$\p=\k_1+\k_2$ which is gained at a parallel escape. In this
configuration, the inter-electron repulsion is strongest and the
angular correlation factor has a relatively narrow width. On the
contrary, the relative motion favors a large relative momentum
$\k=\k_1-\k_2$ which is maximized at an antiparallel escape. In this
configuration, the inter-electron repulsion is weak. This results in a
profoundly large angular correlation width as compared with the
center-of-mass motion.

For simplicity, we consider equal-energy-sharing kinematics
$E_1=E_2$ and restrict ourselves with only the dominant quadrupole
TPDI amplitude. By performing a derivation similar to
Ref.~\cite{KI06}, the quadrupole amplitude can be parametrized as
\ba
\label{quadrupole}
Q&\propto&
\Big[
g_k \{ \hat\k\otimes \hat\k\}_2
+
g_p \{ \hat\p\otimes \hat\p\}_2
\\&&\hs{1cm}+
g_0
\Big\{[\hat\k\times\hat\p]\otimes[\hat\k\times\hat\p]\Big\}_2
\Big]
\cdot
 \{ \e\otimes \e\}_2
\nonumber
\ea
Here we introduced the unit vectors $\hat\p=\p/p$ and  $\hat\k=\k/k$.
The symmetrized amplitudes in \Eref{quadrupole} are defined as
\ba
\label{amplitudes}
g_{k,p}(x)&=&
\sum_{l=0}
\frac12D_l
\big[P''_{l}(x) + P''_{l+2}(x) 
\\ &&\hs{-1.5cm}
\pm2P''_{l+1}(x) \big]
Q_{ll+2}(k_1,k_2)
\mp\frac14C_l P''_{l}(x) Q_{ll}(k_1,k_2),
\nonumber
\\
g_0(x)&=&
\sum_{l=2}
C_l P''_{l}(x) Q_{ll}(k_1,k_2).
\nonumber
\ea
Here the upper and lower sets of signs refers to $g_k$ and $g_p$,
respectively. These amplitudes depend on the interelectron angle
$x=\cos\theta_{12}$ and are expressed via the quadrupole radial matrix
elements $ Q_{ll'}(k_1,k_2)$ and the Legendre polynomial derivatives $
P''_{l}(x)$.  Normalization coefficients $C_l$ and $D_l$ are given in
Ref.~\cite{MMM96}.  Using similar notations, the dipole amplitude of
SPDI \eref{dipole} under the equal energy sharing condition is
parametrized as $D\propto f_p \ (\hat\p\cdot \e)$, $f^+\equiv f_p$,
$f^-=0$. We note that $D$ is {\em linear} with respect to $\hat\p$ and
does not contain $\hat\k$ under the equal energy condition. In
contrast, $Q$ is {\em quadratic} with respect to $\hat\k$ and $\hat\p$
and contains both vectors even when $E_1=E_2$. The amplitude $g_k$
which enters \Eref{quadrupole} with the tensorial product $\{
\hat\k\otimes \hat\k\}_2$ can be associated with the relative motion
of the photoelectron pair described by the vector $\k$. Similarly, the
amplitudes $g_p$ can be associated with the center-of-mass motion and
the amplitude $g_0$ which is entering \Eref{quadrupole} with the
vector product $\hat\k\times\hat\p$ can be associated with the mixed
motion mode.

\begin{figure}[h]
\vs{4cm}
\epsfxsize=6.cm
\epsffile{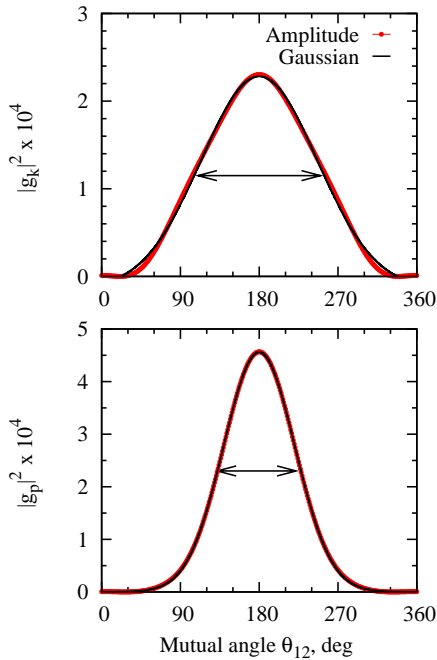}
\caption{
The TPDI amplitudes in the quadrupole channel $g_k$ (top) and $g_p$
(bottom) fitted with the Gaussian ansatz \eref{gauss}. 
The excess energy of 4~eV is shared equally between the photoelectrons
$E_1=E_2=2$~eV. The arrows indicate the Gaussian width parameter 
$\Delta\theta_{12}$. }
\label{fig1}
\end{figure}

To calculate the radial matrix elements $ Q_{ll'}(k_1,k_2)$, we
employed here the same dynamical model as was outlined in our previous
work \cite{KI06}. In this model, the electron-photon interaction was
treated in the lowest-order perturbation theory using the closure
approximation whereas the electron-electron interaction was included
in full using the convergent close-coupling (CCC) method. The model
proved to be capable of describing the angular correlation pattern in
the two-electron continuum in good agreement with non-perturbative,
with respect to the electromagnetic interaction, calculations
\cite{CP02a,HCC05}. 

We calculated amplitudes \eref{amplitudes} in a range of excess
energies $E_1+E_2$ from 1~eV to 20~eV above the double ionization
threshold. We employed a fairly large CCC basis set composed typically
of $25-l$ box-space  target states \cite{BBS03R} with $0\leq l \leq
6$. Convergence of the calculation with respect to the basis size was
thoroughly tested.

\begin{figure}[h]
\bigskip

\epsfxsize=6.5cm
\epsffile{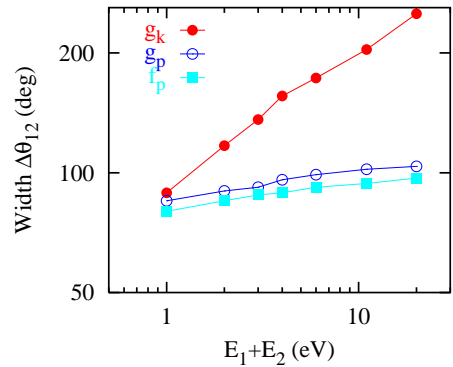}
\caption{
The Gaussian width parameters $\Delta\theta_{12}$ of the amplitudes
$g_k$ (red filled circles), $g_p$ (blue open circles) and $f_p$ (light
blue squares) as functions of the excess energy $E_1+E_2$. 
Extraction of the width
parameters is illustrated  in \Fref{fig1}  for  $E_1=E_2=2$~eV.}
\label{fig2}
\end{figure}

In the whole excess energy range, the amplitude $g_0$ was found
insignificant as compared with $g_{k}$ and $g_{p}$. The latter
amplitudes were fitted with the Gaussian ansatz \eref{gauss}. A
typical quality of the fit can be judged from \Fref{fig1} where the
amplitudes $g_k$ and $g_p$ are exhibited for
$E_1=E_2=2$~eV. The corresponding width parameters $\Delta\theta_{12}$
are plotted in \Fref{fig2}, as a function of energy, along with the width parameter of the dipole amplitude $f_p$.

We observe a much larger Gaussian width of the relative motion
amplitude $g_{k}$ as compared with the center-of-mass motion
amplitudes $g_{p}$ and $f_{p}$, the latter two having a very similar
width. We interpret this stark difference in terms of the strength of
the electron-electron repulsion. This strength is much larger in the
center-of-mass motion which favors large $p$ and hence parallel
emission, as opposed to the relative motion which favors large $k$
associated with anti-parallel emission. We note that the mixed
mode amplitude is typically an order of magnitude smaller than
the pure mode amplitudes  $g_{k}$ and  $g_{p}$ thus supporting our
notion of distinct escape modes. 

\begin{figure}[h]
\vs{13cm}
\epsfxsize=6.cm
\epsffile{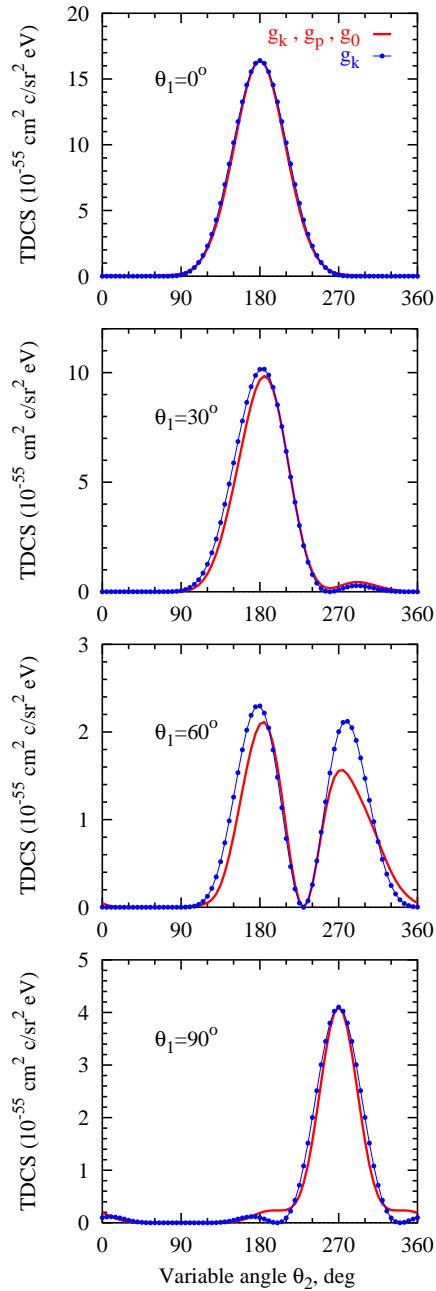}
\caption{
Triply-differential cross-section
$d\sigma/dE_1d\Omega_1d\Omega_2\propto |Q|^2$ of the TPDI of He at
$E_1=E_2=2$~eV in the coplanar kinematics is plotted as a function of
the variable escape angle $\theta_2$. Various panels from top to
bottom correspond to fixed escape angles $\theta_1=0^\circ$,
$30^\circ$, $60^\circ$ and $90^\circ$. The red solid line corresponds
to the full quadrupole amplitude \eref{coplanar} whereas the blue
dotted line exhibits the contribution of the $g_k$ term only.}
\label{fig3}
\vs{-1cm}
\end{figure}

In the case of co-planar geometry when all three vectors $\k_1$,
$\k_2$ and $\e$ belong to the same plane, we can choose the coordinate
frame as   $z\ \| \ \k_1$ and $x \ \| \ [\k_1\times\k_2] $. In this
case, \Eref{dipole} and \Eref{quadrupole} are simplified to
\ba
D&\propto& f_p \ (\cos\theta_1+\cos\theta_2)
\\
\nonumber
Q&\propto&
g_k
\Bigg[
(\cos\theta_1-\cos\theta_2)^2-
\frac23 (1-\cos\theta_{12})
\Bigg]
\\&+&
g_p
\Bigg[
(\cos\theta_1+\cos\theta_2)^2-
\frac23 (1+\cos\theta_{12})
\Bigg]
\nonumber
\\&+&
\frac13
g_0
(\cos^2\theta_{12}-1)
\label{coplanar}
\ea
From these equations, it is seen that the amplitudes $f_p$ and $g_p$
related to the center-of-mass mode, and the amplitude $g_k$ related to
the relative motion mode, contribute quite differently to the
corresponding matrix elements. All three amplitudes peak strongly
near $\theta_{12}=180^\circ$. However, the kinematic factors
corresponding to $f_p$ and $g_p$ have nodes at this angle whereas the
kinematic factor accompanying $g_k$ has a peak. As a result, the
term proportional to $g_k$ dominates strongly the quadrupole
amplitude. This dominance is illustrated in \Fref{fig3} where the
triply-differential cross-section
$d\sigma/dE_1d\Omega_1d\Omega_2\propto |Q|^2$ of the TPDI of He at
$E_1=E_2=2$~eV and the coplanar kinematics is plotted as a function of
the variable escape angle $\theta_2$. Various panels from top to
bottom correspond to fixed escape angles $\theta_1=0^\circ$,
$30^\circ$, $60^\circ$ and $90^\circ$. Two calculations are displayed
in the figure: one with the full quadrupole amplitude \Eref{coplanar}
and another with only the $g_k$ contribution. The dominance of the $g_k$
term is evident at all fixed angles $\theta_2$.  It is particularly
strong at a fix escape angle $\theta_1=0$ when nearly all the
contribution to the quadrupole amplitude comes from the $g_k$ term.

We note that in our previous paper \cite{KI06} we introduced a
slightly different set of amplitudes in the quadrupole channel:
\ba
\label{parametrization1}
Q&\propto&
\frac23
\Bigg\{
g^+
\Big[P_2(\cos\theta_1)+P_2(\cos\theta_2)\Big]
\\&&\hs{4mm}
\frac12
g_s
\Big[3\cos\theta_1\cos\theta_2 - \cos\theta_{12}\Big]
\nonumber
\\&&\hs{4mm}
\frac12
g_0
(\cos^2\theta_{12}-1)
\Bigg\}
\nonumber
\ea
where 
\ba
\label{amplitudes1}
g^+&=& \ \  g_p+g_k\\
g_s&=&2(g_p-g_k)
\nonumber
\ea


Although the set of amplitudes \eref{amplitudes1} gives an identical
quadrupole amplitude, it does not provide such a clear separation of
the center-of-mass and relative motion modes. Our earlier attempt in
Ref.~\cite{KI06} to apply the Gaussian ansatz \eref{gauss} showed no
such a clear systematic behavior of the angular correlation width with
respect of the excess energy as exhibited in \Fref{fig2}.

\newpage
~~~~~~~~~~~~
\newpage
In conclusion, we demonstrated the presense of the two distinct
photoelectron escape modes following two-photon double ionization of
He. One of the modes corresponds to the center-of-mass motion of the
photoelectron pair. It enhances the total momentum of the pair and
therefore favors the parallel emission. The inter-electron repulsion
is strong in this mode and the angular correlation width is relatively
small. The other, relative motion, mode enhances the relative momentum
of the pair and therefore favors the antiparallel emission. The
inter-electron repulsion is much weaker in this mode and the angular
correlation width is march larger. Both modes are fully symmetric
and present under the equal energy sharing condition. In contrast, the
single-photon double ionization has only one fully symmetric mode
which is associated with the center-of-mass motion.  This mode, in
terms of the angular correlation width, is very similar to the
center-of-mass motion mode in two-photon double ionization. The
presense of two modes is simply a reflection of the quadratic
tensorial structure of the quadrupole photoionization amplitude as
compared to the linear structure of the dipole photoionization
amplitude.

The authors wish to thank Australian Partnership for
Advanced Computing (APAC) and ISA Technologies, Perth, Western
Australia, for provision of their computing facilities.  Support of
the Australian Research Council in the form of Discovery grant
DP0451211 is acknowledged.


\begin{thebibliography}{10}

\bibitem{URD03}
J. Ullrich {\it et~al.}, Rep.~Prog.~Phys. {\bf 66},  1463  (2003).

\bibitem{BS00}
J.~S. Briggs and V. Schmidt, J.~Phys.~B {\bf 33},  R1  (2000).

\bibitem{KA00}
G.~C. King and L. Avaldi, J.~Phys.~B {\bf 33},  R215  (2000).

\bibitem{AH05}
L. Avaldi and A. Huetz, J. Phys. B {\bf 38},  S861  (2005).

\bibitem{HSWM91}
A. Huetz, P. Selles, D. Waymel, and J. Mazeau, J.~Phys.~B {\bf 24},  1917
  (1991).

\bibitem{MSH97}
L. Malegat, P. Selles, and A. Huetz, J.~Phys.~B {\bf 30},  251  (1997).

\bibitem{S95}
S.~J. Schaphorst {\it et~al.}, J. Electron Spectrosc. Relat. Phenom. {\bf 76},
  229  (1995).

\bibitem{Rau76}
A.~R.~P. Rau, J.~Phys.~B {\bf 9},  L283  (1976).

\bibitem{Feagin84}
J.~M. Feagin, J.~Phys.~B {\bf 17},  2433  (1984).

\bibitem{KB02}
A.~S. Kheifets and I. Bray, Phys.~Rev.~A {\bf 65},  022708  (2002).

\bibitem{KBB03}
P. Bolognesi {\it et~al.}, J.~Phys.~B {\bf 36},  L241  (2003).

\bibitem{KKKB05}
A. Knapp {\it et~al.}, J.~Phys.~B {\bf 28},  645  (2005).

\bibitem{KI06}
A.~S. Kheifets and I.~A. Ivanov, J. Phys. B {\bf 39},  1731  (2006).

\bibitem{DAMOP06}
E.~A. Pronin {\it et~al.}, {\em 37th Meeting of the Division of Atomic,
  Molecular and Optical Physics} (American Physical Society, Knoxville, TN,
  2006).

\bibitem{MMM96}
N.~L. Manakov, S.~I. Marmo, and A.~V. Meremianin, J.~Phys.~B {\bf 29},  2711
  (1996).

\bibitem{CP02a}
J. Colgan and M.~S. Pindzola, Phys. Rev. Lett. {\bf 88},  173002  (2002).

\bibitem{HCC05}
S.~X. Hu, J. Colgan, and L.~A. Collins, J.~Phys.~B {\bf 38},  L35  (2005).

\bibitem{BBS03R}
I. Bray, K. Bartschat, and A.~T. Stelbovics, Phys.~Rev.~A {\bf 67},  060704(R)
  (2003).

\end{thebibliography}

\end{document}